\documentstyle[epsfig]{aipproc}

\begin{document}
\title{Top Quark Yukawa Couplings and New Physics}

\author{S. Dawson$^*$ and L. Reina$^WY{\dagger}$}
\address{$^*$Physics Department, Brookhaven National
Laboratory\\
Upton, NY  11973 \\
$^{\dagger}$Physics Department, Florida State University\\
 Tallahassee, FL  32306 }
\newcommand{\SLASH}[2]{\makebox[#2ex][l]{$#1$}/}
\newcommand{\Dslash}{\SLASH{D}{.5}\,}
\newcommand{\dslash}{\SLASH{\dd}{.15}}
\newcommand{\kslash}{\SLASH{k}{.15}}
\newcommand{\pslash}{\SLASH{p}{.2}}
\newcommand{\qslash}{\SLASH{q}{.08}}

\maketitle

\begin{abstract}

We discuss  associated production of a
Higgs boson with a pair of $t {\overline t}$ quarks at a
future high energy $e^+e^-$ collider.
The process $e^+e^-\rightarrow t {\overline t}h$
 is particularly sensitive
to the presence of new physics and we consider the MSSM and 
models with extra dimensions at the $TeV$ scale as examples.

\end{abstract}

\section*{Introduction}

The present and next generation of colliders will help elucidate the
nature of the electroweak symmetry breaking and to determine
if the symmetry  breaking is due to the presence of a Higgs boson.
  Precision fits of the Standard
Model seem to indirectly point at the existence of a light Higgs boson
($M_{h}<170-210$ GeV)\cite{osaka}, while the Minimal
Supersymmetric Standard Model (MSSM) requires the existence of a
scalar Higgs lighter than about 130 GeV. Thus, 
a Higgs discovery by the Tevatron or the LHC
in the mass range around $120-130$ GeV seems 
a likely possibility. The role
 of the next generation of $e^+e^-$ colliders will
therefore not be to make a Higgs discovery, but
rather   to obtain precision measurements
of the Higgs properties and to
 use them to understand the underlying mechanism
of electroweak symmetry breaking.
 
The production of a Higgs boson  in
association with a pair of $t {\overline t}$ quarks is of
particular interest
for two reasons. First, the $t\bar t h$ production mode can be an
important discovery  mode for a Higgs boson around $120-130$ GeV at the
LHC\cite{atlasreport} or the Tevatron\cite{davidetal}. 
Second,  this  channel offers a direct handle
on the Yukawa coupling of the top quark to the Higgs boson.

\begin{figure}[t]
\centering
\epsfysize=2.2in
\leavevmode\epsffile{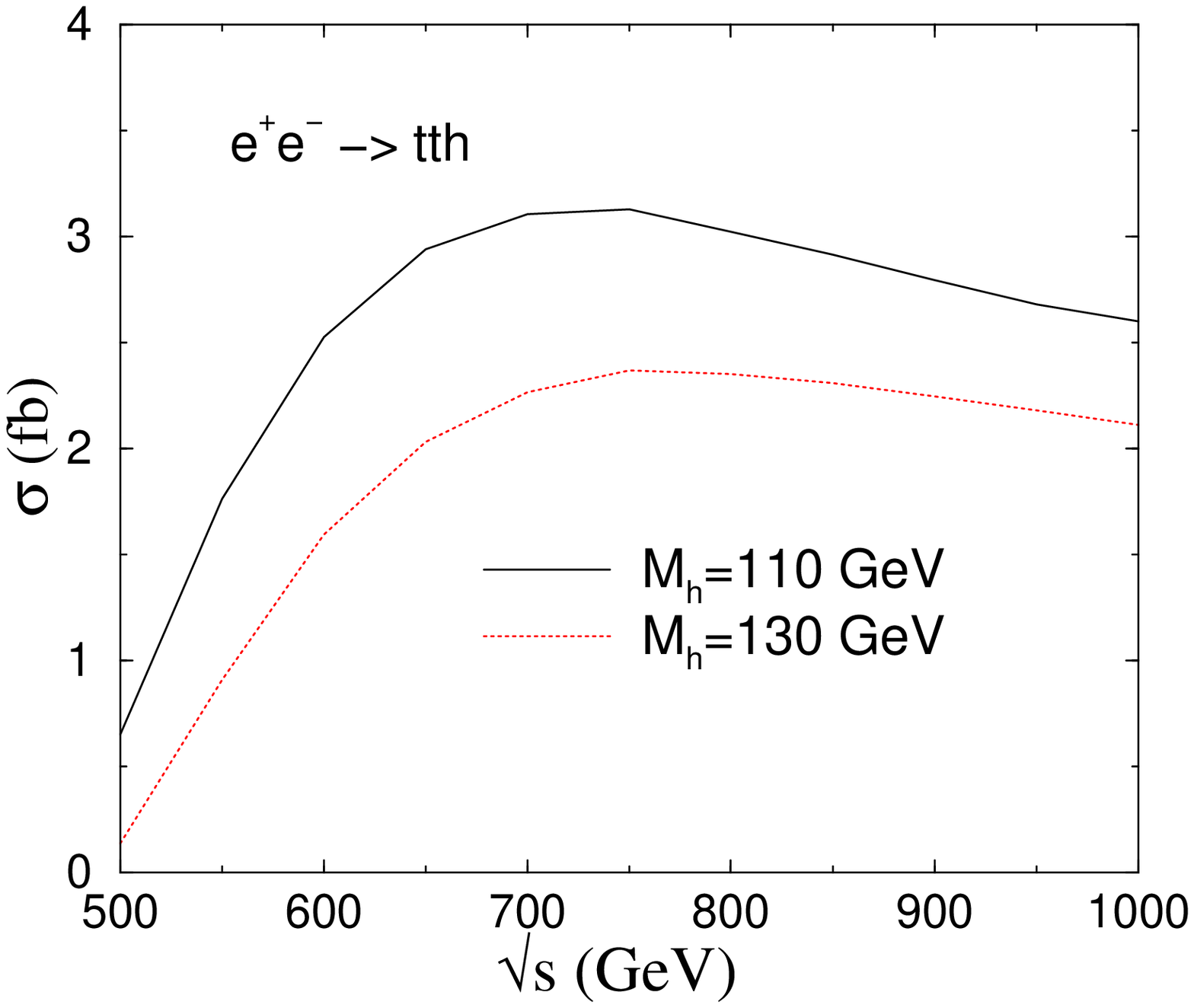}
\hspace{1.truecm}
\epsfysize=2.2in
\leavevmode\epsffile{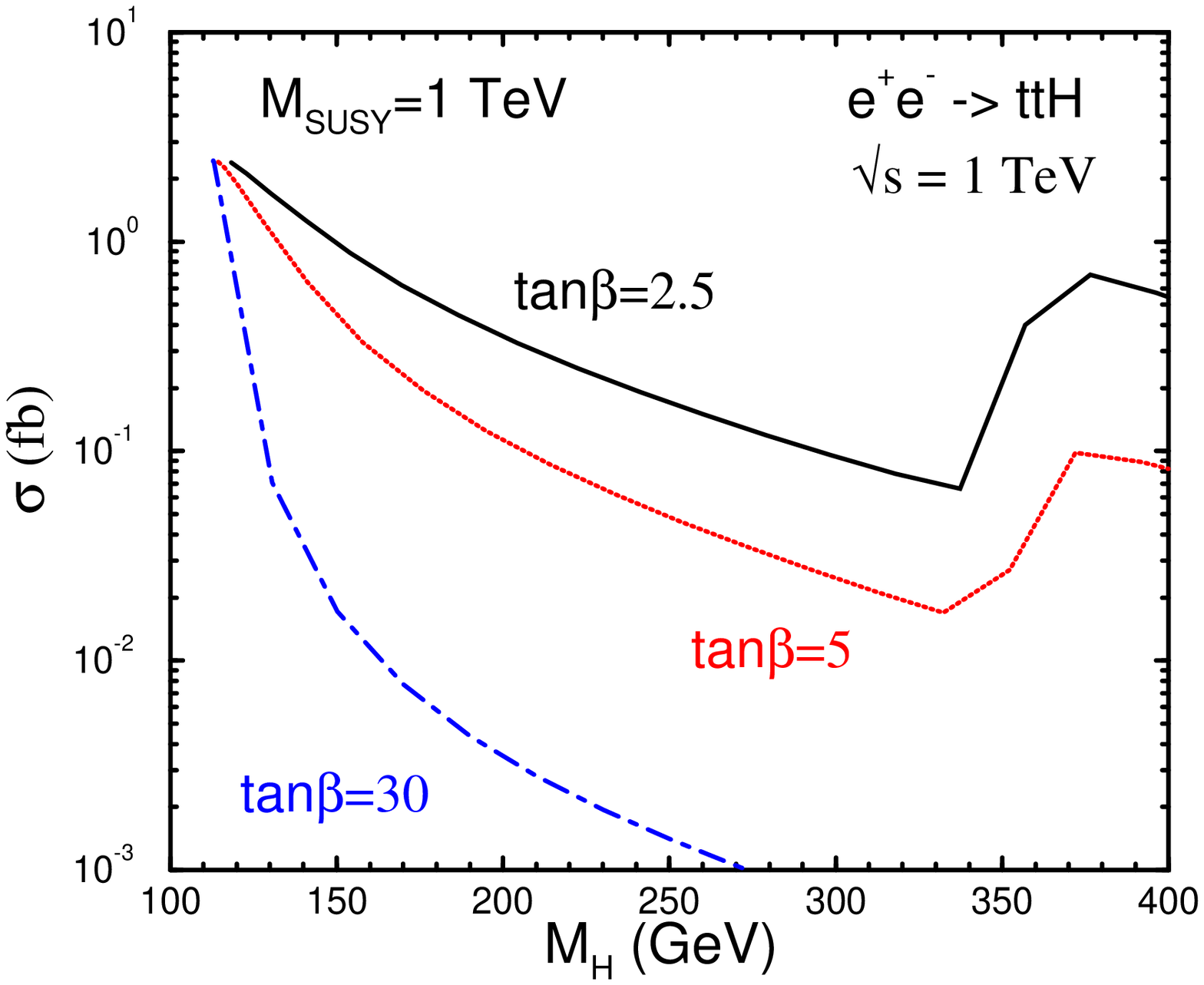}
\caption[]{\emph{Left plot}: Standard Model cross
section for $t {\overline t}h$
production as a function of the center of mass energy, $\sqrt{s}$.
\emph{Right plot}: Cross section for $e^+e^-\rightarrow t 
{\overline t}H$, where $H$ is the heavier of the neutral
Higgs bosons of the MSSM.}
\label{fig:lhc}
\end{figure}


\section{Associated top-Higgs production at a high energy 
$e^+e^-$ collider}

The process $e^+e^-\rightarrow t\bar t h$ has been studied 
extensively
and  has been calculated
both in the SM and in the MSSM at
$O(\alpha_s)$\cite{us,zerwasetal}. For a SM Higgs, the factor
$K=\sigma_{NLO}/\sigma_{LO}$ at $\sqrt{s}\!=\!500$ GeV is in the range
$(1.4-2.4)$, depending on $M_h$. However the cross section is
drastically suppressed by phase space and for $M_h\!\simeq\!120-130$
GeV is of the order of $0.1\,fb$. On the other hand, at
$\sqrt{s}\!=\!1$ TeV the cross section for $M_h\!=\!120-130$ GeV is 
about $2\, fb$ and is only slightly reduced by QCD corrections
($K\!=\!0.8-0.9$).  
The energy dependence of the rate  is shown in Fig. 1 and it is clear
that the optimal energy is somewhere around $\sqrt{s}\sim 700-800~GeV$.

Although the cross section is  small, the signature for $t\bar t
h$ production is spectacular. The possibility of fully reconstructing
the two top quarks in the final state allows efficient discrimination of
the signal over the background, and together with a good b-tagging
efficiency is crucial for increasing the precision with which the top
Yukawa coupling can be measured.  A detailed
simulation\cite{juste} of $e^+e^-\rightarrow t\bar th$ for a $120~GeV$
 Higgs,
at $\sqrt{s}\!=\!800$ GeV, found that the top Yukawa coupling 
could potentially
be measured with a precision of 5.5\% when optimal b-tagging
efficiency is assumed. At lower energies, and 
for heavier Higgs masses, the precision
deteriorates rapidly.\cite{us_analysis}

\section{MSSM}

Associated production of a Higgs boson and a $ t {\overline t}$
pair in the MSSM can be very different from  the Standard Model since
the $t 
{\overline t}h$ coupling can be significantly suppressed for small $M_A$,
(where $h$ is now the lightest neutral Higgs boson of the MSSM).
  In addition, the presence of additional Higgs
bosons leads to the possibility  of new processes and
 of resonance effects.  An example
is $e^+e^-\rightarrow AH\rightarrow t {\overline t} H$, which is illustrated
in the right hand plot of Fig. 1. ($H$ is the heavier of the neutral
Higgs bosons of the MSSM).
The resonance at $2 M_t$ is apparent.
 The
main channels in the MSSM
 are the scalar ones, i.e. $t\bar th$ and $t\bar tH$,
since the pseudoscalar mode $t\bar tA$ is very suppressed over most
of the MSSM parameter space.  Over most of the $tan\beta-M_A$ plane,
it is possible to observe either $t\bar th$ or  $t\bar tH$ with
a rate greater than $1~fb$ at $\sqrt{s}=500~GeV$.\cite{us}

\section{Models with extra dimensions at the TEV scale}

Scenarios such as that proposed by Arkani-Hamed, Dimopoulos, and 
Dvali,\cite{add}  (ADD) in which gravity propagates in $d+4$
dimensions can have a dramatic effect on Higgs physics.  In such
models the gravitational interactions  occur at a scale $M_S$, which
can be as small as a $TeV$.  The extra $d$ dimensions are then
compactified on a torus of radius $R$.  This leads to a relationship
between  Newton's Constant, $G_N$, and the other scales,
\begin{equation}
{1\over G_N}\sim R^d M_S^{d+2} \quad .
\end{equation}
These models have a tower of very light Kaluza Klein excitations which
couple weakly to ordinary matter ($\sim 1/M_{pl}$).
  However, the density of the light
Kaluza Klein modes is high (for large $R$) and so the collective
interactions build up to electroweak strength.
At tree level, the contributions of the spin-$0$ Kaluza Klein modes to $e^+e^-
\rightarrow t {\overline t}H$ are suppressed by $m_e$, so only the
spin $2$ states are relevant here.

It is  straightforward
 to find the effective $4-$fermion and $f {\overline f}
hh$ interactions due to the graviton exchanges which contribute
to $e^+e^-\rightarrow t {\overline t}h$.\cite{han}
  For 
$f(p)+{\overline f}(q) \rightarrow f^\prime(p^\prime) +
 {\overline f}^\prime(q^\prime)$,
we have
\begin{eqnarray}
{\cal A}_{ffff}&=&
{C(s)\over 16}\biggl\{ {\overline v}(q) 
(\SLASH{p^\prime}{.2}-\SLASH{q^\prime}{.2}) u(p)
	{\overline v}(p^\prime)(\SLASH{p}{.2}-
\SLASH{q}{.2}) v(q^\prime)
\nonumber \\
&&+(p-q)\cdot (p^\prime-q^\prime){\overline u}(q) \gamma^\mu u(p)
	{\overline u}(p^\prime)\gamma_\mu v(q^\prime)\biggr\}~~,
\end{eqnarray}
while for 
$f(p)+{\overline f}(q) \rightarrow h(p_{h1}) + h(p_{h2})$,
the effective interaction is,
\begin{equation}
{\cal A}_{ffhh}=
 {C(s)\over 2}  (p-q)\cdot p_{h_1} {\overline v}(q) 
\SLASH{p_{h_1}}{.2}~~ u(p)~~.  
\end{equation}
The constant $C(s)$ represents the effect of the sum over all the
Kaluza Klein modes (with masses $M_i$) and is given by\cite{han}
\begin{equation}
C(s)=
-16\pi G_N\sum_i {1\over s-M_i^2}\rightarrow_{(\sqrt{s}<<M_S, d>2)}
 {16\pi\over d-2}\biggl({1\over M_S}\biggr)^4~~.
\end{equation}

Calculating the complete rate
for $e^+e^-\rightarrow t {\overline t} h$, including the 
Standard Model interactions, the new interactions of Eqs. (2) and (3),
and the relevant interference terms, gives
 the
rates  in the ADD scenario  shown
in Fig. 2 for $d=3$ extra dimensions.
  At $\sqrt{s}=1~TeV$, the rate can be enhanced by several
orders of magnitude if the gravitational interactions are at the $1-2~TeV$
scale.  The energy distribution of the Higgs boson, $x_h={2 E_h/\sqrt{s}}$,
is slightly shifted to lower values of $x_h$ by the presence of the
gravitational interactions.

\section{Conclusion}

The process $e^+e^- \rightarrow t {\overline t} h$
 provides a sensitive probe of new physics effects.  It is best measured
at $\sqrt{s}=700-800~GeV$ and requires the highest possible luminosity.
\begin{figure}[t]
\centering
\epsfysize=2.2in
\leavevmode\epsffile{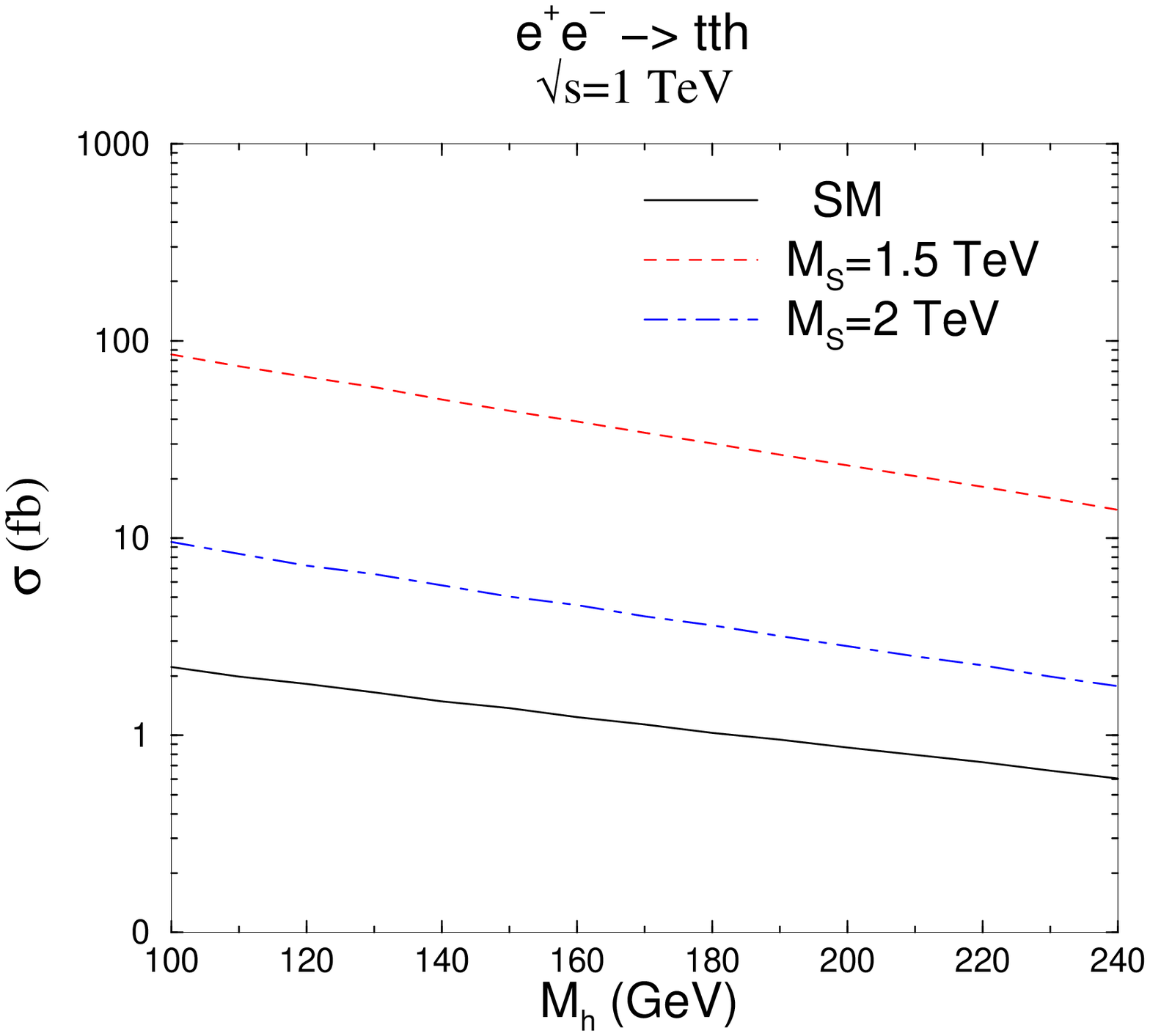}
\hspace{1.truecm}
\epsfysize=2.2in
\leavevmode\epsffile{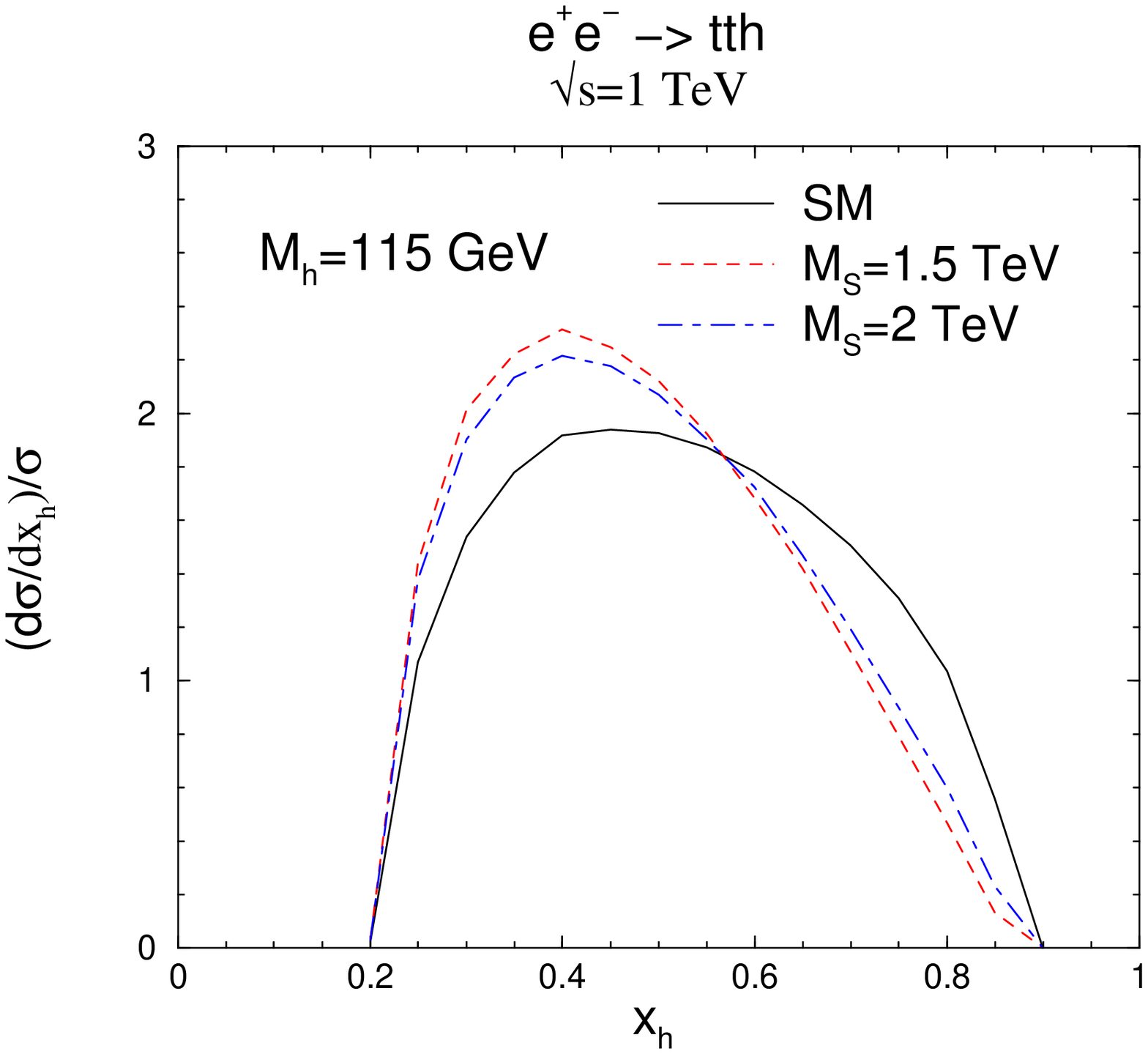}
\caption[]{\emph{Left plot}: Cross
section for $t\bar th$ production in the ADD model with $d=3$
extra dimensions and where gravitational interactions
occur at the scale $M_S$.
\emph{Right plot}: Energy distribution of the Higgs boson in the ADD model.}
\label{fig:lhc}
\end{figure}

\end{document}